\documentclass[prl,aps,showpacs,superscriptaddress,floatfix]{revtex4}
\usepackage[T1]{fontenc}
\usepackage[francais]{layout}
\usepackage{graphics}
\usepackage{amssymb,amsmath}
\newcommand{\ket}[1]{|#1\rangle}
\newcommand{\bra}[1]{\langle#1|}

\newcommand{\one}{\leavevmode\hbox{\small1\normalsize\kern-.33em1}}

\begin{document}
\title{A Photonic Quantum Information Interface}

\author{S. Tanzilli}
\email{sebastien.tanzilli@physics.unige.ch}
\affiliation{Group of Applied Physics, University of Geneva, 20 rue de l'École de Médecine, CH-1211 Geneva 4, Switzerland}
\author{W. Tittel}
\affiliation{Group of Applied Physics, University of Geneva, 20 rue de l'École de Médecine, CH-1211 Geneva 4, Switzerland}
\author{M. Halder}
\affiliation{Group of Applied Physics, University of Geneva, 20 rue de l'École de Médecine, CH-1211 Geneva 4, Switzerland}
\author{O. Alibart}
\affiliation{Laboratoire de Physique de la Matière Condensée, Université de Nice--Sophia Antipolis, Parc Valrose, 06108 Nice Cedex 2, France}
\author{P. Baldi}
\affiliation{Laboratoire de Physique de la Matière Condensée, Université de Nice--Sophia Antipolis, Parc Valrose, 06108 Nice Cedex 2, France}
\author{Nicolas Gisin}
\affiliation{Group of Applied Physics, University of Geneva, 20 rue de l'École de Médecine, CH-1211 Geneva 4, Switzerland}
\author{Hugo Zbinden}
\affiliation{Group of Applied Physics, University of Geneva, 20 rue de l'École de Médecine, CH-1211 Geneva 4, Switzerland}

\begin{abstract}
{\bf Quantum communication is the art of transferring quantum
states\cite{Tittel-Weihs}, or quantum bits of information
(qubits), from one place to another. On the fundamental side, this
allows one to distribute entanglement and demonstrate quantum
nonlocality over significant
distances\cite{Tittel98,MarcikicCrypto50,Weihs-NL,Resch,Peng}. On
the more applied side, quantum cryptography offers, for the first
time in human history, a provably secure way to establish a
confidential key between distant partners\cite{GAP-RMP02}. Photons
represent the natural flying qubit carriers for quantum
communication, and the presence of telecom optical fibres makes
the wavelengths of 1310 and 1550\,$nm$ particulary suitable for
distribution over long distances. However, to store and process
quantum information, qubits could be encoded into alkaline atoms
that absorb and emit at around 800\,$nm$
wavelength\cite{Monroe,Shapiro}. Hence, future quantum information
networks made of telecom channels and alkaline memories will
demand interfaces able to achieve qubit transfers between these
useful wavelengths while preserving quantum coherence and
entanglement\cite{Shapiro,Kumar,DeMartini}. Here we report on a
qubit transfer between photons at 1310 and 710\,$nm$ via a
nonlinear up-conversion process with a success probability greater
than 5\%. In the event of a successful qubit transfer, we observe
strong two-photon interference between the 710\,$nm$ photon and a
third photon at 1550\,$nm$, initially entangled with the
1310\,$nm$ photon, although they never directly interacted. The
corresponding fidelity is higher than 98\%.}
\end{abstract}


\maketitle

%
%


Superposition of quantum states and entanglement are the
fundamental resources of quantum communication and quantum
information processing\cite{Tittel-Weihs}. From an abstract point
of view, the nature of the carrying particle is irrelevant since
only amplitudes and relative phases are exploited to encode the
elementary quantum bits of information (qubits). Historically,
experiments have proven many times that the fascinating properties
of quantum correlations can be observed with pairs of
photons\cite{Tittel-Weihs}, trapped
ions\cite{RoweIons01,SchmidtIons03,Wineland-Gate}, trapped
atoms\cite{MandelAtom03}, or cold gases\cite{JulsgaardGas03}.
However, the most appropriate carrier and associated encoding
observable depend on the specific task. Photons have been proven
suitable to transmit quantum
information\cite{Tittel-Weihs,Tittel98,MarcikicCrypto50,Weihs-NL,Resch,Peng,GAP-RMP02},
and atoms or ions to store\cite{Polsik} and
process\cite{SchmidtIons03,Wineland-Gate} it. Photonic
entanglement often relies on
polarization\cite{Weihs-NL,Peng,Resch,Aspect,Kwiat99,Kukle},
energy-time\cite{Tittel98}, or time-bin\cite{MarcikicCrypto50}
coding. Depending on the quantum communication channel, the
wavelength of the photonic carrier is also important. The use of
telecom wavelengths (1310 and 1550\,$nm$) is particularly
advantageous when employing optical
fibres\cite{Tittel98,MarcikicCrypto50} while free space
transmission is mostly based on shorter wavelengths\cite{Resch,Peng}.\\
\indent Future realizations of quantum networks, containing
elementary quantum processors and memories, connected by
communication channels, require quantum interfaces capable of
transferring qubit states from one type of carrier to another.
This demands the reversible mapping between photons and atoms
which also includes the mapping between photons of different
wavelengths\cite{Shapiro}. However, as opposed to the reproduction
of classical information between different media, it is not
possible to merely measure the properties of a given quantum
system and replicate them accordingly, as a result of the
no-cloning theorem\cite{Wootters}. Nevertheless it is possible to
resort to a transfer of the quantum information based on an
interaction that maintains the coherence properties of the initial
quantum system.\\
\indent In this letter, we demonstrate a direct quantum interface
for photonic qubits at different wavelengths.
As illustrated in figure 1, an arbitrary qubit
carried by a flying telecom photon at
1310\,$nm$ ($\lambda_B$), initially entangled with a photon at
1550\,$nm$ ($\lambda_A$), is coherently transferred to another
photon at a wavelength of 710\,$nm$ ($\lambda_{B'}$) using sum
frequency generation (SFG). The final wavelength is
close to that of alkaline atomic transitions.\\
\indent
In the following, we first present the theoretical description of our
quantum interface. We then recall how to create and characterize
maximally energy-time entangled photon-pairs. The coherent quantum information
transfer, taking advantage of an up-conversion stage, is then
introduced. Finally, we show how this operation preserves the
initial entanglement by measuring the two-photon interference
between the 710 and 1550\,$nm$ photons, and thus we demonstrate a
universal qubit transfer.

%
\vspace{0.5cm}
\indent Consider three systems labelled A (held by Alice), B and
B' (held by Bob). In our case these systems are modes of the
electro-magnetic field in optical fibers. Initially, B' is in the
vacuum state denoted $\ket{0}$, while A and B may be entangled,
with Schmidt coefficients $c_1$ and $c_2$:
\begin{equation}
\ket{\Psi}_{in} = \big( c_1\ket{\alpha_1}_A \otimes
\ket{\beta_1}_B + c_2\ket{\alpha_2}_A \otimes \ket{\beta_2}_B
\big) \otimes \ket{0}_{B'} \label{initial-state}
\end{equation}
The pairs of orthogonal states $\ket{\alpha_j}$ and
$\ket{\beta_j}$, $j \in \left \{ 1,2 \right \}$, span Alice and
Bob's qubit space, respectively. For example, $\ket{\alpha_j}$
could represent one photon in either the vertical (j=1) or
horizontal (j=2) polarization state, or, as in our case, one
photon in the first (j=1) or second (j=2) time-bin state. The
desired state after a successful transfer from B to B' reads:
\begin{equation}
\ket{\Psi}_{transfer} = \ket{0}_B\otimes \big( c_1\ket{\alpha_1}_A
\otimes \ket{\beta_1}_{B'} + c_2\ket{\alpha_2}_A \otimes
\ket{\beta_2}_{B'} \big) \label{transfer-state}
\end{equation}
Such a transfer is achieved by an interaction described by the
effective Hamiltonian
\begin{eqnarray}
\mathcal{H} &=& \one_A \otimes \Big( g_1 \ket{0}_B\bra{\beta_1}
\otimes \ket{\beta_1}_{B'}\bra{0}\, +\, g_2 \ket{0}_B\bra{\beta_2}
\otimes \ket{\beta_2}_{B'}\bra{0} \Big) + H.c. \label{H}
\end{eqnarray}
where $g_1$ and $g_2$ are coupling constants. The evolution can
easily be computed (assuming an interaction time of 1 unit):
\begin{eqnarray}
e^{-i\mathcal{H}}\ket{\Psi}_{in}&=&\cos(|g_1|)c_1\ket{\alpha_1}_A
\otimes \ket{\beta_1}_B \otimes \ket{0}_{B'} -
i\frac{\sin(|g_1|)}{|g_1|} g_1 c_1 \ket{\alpha_1}_A \otimes \ket{0}_B \otimes \ket{\beta_1}_{B'} \label{1stLine} \\
&+&\cos(|g_2|)c_2\ket{\alpha_2}_A \otimes \ket{\beta_2}_B \otimes
\ket{0}_{B'} - i\frac{\sin(|g_2|)}{|g_2|} g_2 c_2 \ket{\alpha_2}_A
\otimes \ket{0}_B \otimes \ket{\beta_2}_{B'} \label{2ndLine}
\end{eqnarray}
Note that in the second terms in (\ref{1stLine}) and
(\ref{2ndLine}), the coefficients $c_j$ are multiplied by $g_j$.
This implies that the transfer preserves the quantum coherence,
i.e. is a Quantum Information (QI) transfer, if and only if the
two coupling constants are equal, both in amplitude and phase:
$g_1=g_2\equiv g$. In other words, the transfer can be achieved
with perfect fidelity provided this condition is satisfied. In
such a case the evolution simplifies to:
\begin{equation}
e^{-i\mathcal{H}}\ket{\Psi}_{in}=\cos(|g|)\ket{\Psi}_{in}-i
g\frac{\sin(|g|)}{|g|}\ket{\Psi}_{transfer}
\label{general-condition}
\end{equation}
Consequently, the transfer probability is 1 for $|g|=\pi/2$.
%

%
\vspace{0.5cm}
Figure 2 presents our 2-photon source $S$. It consists of a CW
pump laser diode with a coherence length $\mathcal{L}_p^c$
exceeding 300\,$m$, attenuated down to a few $\mu W$ at
$\lambda_p$ = 711.6\,$nm$, and of a quasi-phase matched
periodically poled lithium niobate waveguide\cite{Tanz02,Bana01}
($PPLN/W1$). The poling period is chosen so as to create
down-converted pairs of photons whose wavelengths are
centered at 1312 and 1555\,$nm$, respectively.
After separation at a fiber-optic wavelength
division multiplexer (WDM), the 1555 and 1312\,$nm$ photons are
sent to Alice and Bob, respectively, using standard telecom optical fibres.\\
\indent The long coherence length of the pump laser implies that
each pump photon has a very well defined energy. Accordingly, the
down-conversion process yields pairs of photons that are energy
correlated as governed by energy conservation: each photon from a
pair has an uncertain energy, uncertain in the usual quantum
sense, but the sum of the energies of the two photons from a pair
is very well defined. In addition, the two photons are also time
correlated, since they are emitted simultaneously within their
coherence time. However, the emission time of a pair is uncertain
within the pump laser coherence. Hence the paired photons are
entangled à la EPR\cite{Aspect-Nature99}, energy and time replacing
the historical position and momentum variables, respectively. More
precisely, the processes of photon-pair creation at different
times separated by $\Delta \tau$ are coherent as long as $\Delta
\tau << \mathcal{L}_p^c / c$, where $\mathcal{L}_p^c$ is the pump
laser's coherence length. The underlying quantum coherence
of the photon-pairs therefore comes from the pump laser itself.\\
\indent In order to reveal and characterize the coherence carried
by these energy-time entangled photon-pairs we perform a
Franson-type experiment\cite{Franson89} and infer the degree of
entanglement from the measured two-photon interference visibility.
For this purpose the two photons are sent to two analyzers, one on
Alice's and the other on Bob's side, that consist, in our case, of
unbalanced Michelson interferometers made from standard telecom
fibres, fiber-optic beam-splitters, and Faraday
mirrors\cite{Tittel-Weihs} as depicted in figure 2. To prevent
single photon interference, the optical path length difference in
each interferometer ($\Delta L_A$ and $\Delta L_B$, respectively)
has to be much greater than the coherence length of the single
photons, $\mathcal{L}_{s,i}^c \simeq 150\,\mu m$, deduced from the
15\,$nm$ spectral bandwidth of the down-converted light. Each
single photon therefore has an equal probability of $\frac{1}{2}$
to exit at one of the outputs of its analyzer. Moreover, to
maximize two-photon interference, these analyzers have to be
equally unbalanced, $\Delta L_A \simeq \Delta L_B = \Delta L$,
within the coherence length of the single photons. At the same
time, however, since an entangled pair represents the quantum
object to be analyzed, both $\Delta L$ ($\sim 20\,cm$ in our case)
have to be much smaller than the coherence length of the pair
which is given
by the pump laser ($\mathcal{L}_{p}^c > 300\,m$).\\
\indent
At the output of each interferometer, the single photons are
detected.
Using a time-resolved coincidence detection between Alice and Bob
as depicted in figure 2, one can post-project the initial energy-time
entangled state onto a time-bin entangled state of the form
\begin{equation}
\ket{\Psi}_{post} = \frac{1}{\sqrt{2}} \left( \ket{s_A,s_B} + e^{i\left( \phi_A + \phi_B \right)} \ket{l_A,l_B}  \right)
\label{entanglement}
\end{equation}
where $\phi_A$ and $\phi_B$ are phases associated with the path
length difference $\Delta L_A$ and $\Delta L_B$ of the related interferometers.
The state (\ref{entanglement}) corresponds to the events where the
down-converted photons both travel through the same arms of the
interferometers, i.e. $s_A,s_B$ or $l_A,l_B$, these two possibilities being
undistinguishable. Thus, varying the combined phase ($\phi_A + \phi_B$),
we observe sinusoidal interference fringes in the coincidence rate with net
(accidental coincidences discarded) and raw visibilities of
97.0$\pm 1.1$\% and 87.4$\pm 1.1$\%, respectively (see Fig. 2).
Because almost all accidental coincidences are due to detector
dark counts, the purity of the created state, and hence the
performance of the source, should be characterized by the net
visibility. Since this figure of merit is very close to the 100\%
predicted by quantum theory, we conclude that our source provides
a state close to a pure maximally entangled state, in particular
suitable to violate the Bell inequalities\cite{Aspect-Nature99}. Note that
the missing 3\% in the net visibility is most likely due to the measurement
technique rather than the state preparation\cite{Tittel-Weihs}.

%
\vspace{0.5cm}
Now that we have characterized our entanglement resource, we
proceed to the quantum information transfer, i.e. the transfer of
the qubit carried by Bob's particle to another photon of shorter
wavelength. To this end, we replace the fiber-optic interferometer
and the Ge-APD on Bob's side by another nonlinear PPLN waveguide
($PPLN/W2$), a bulk-optic Michelson interferometer, and a Silicon
avalanche photodiode (Si-APD, {\it D}$_B$) as depicted in figure 3.
This second nonlinear crystal, together with a power reservoir
(PR) consisting of a CW coherent laser at $\lambda_{PR} =
1560\,nm$ and an erbium-doped fiber amplifier (EDFA), serves as an
up-conversion stage for the incoming 1312\,$nm$ photons produced
by the previously described source $S$.
As a result of this operation, which is the
opposite process to down-conversion, photons at 1312 and
1560\,$nm$ are annihilated and photons at 712.4\,$nm$ are created
on Bob's side. The coherence of the PR directly relates to
the phase difference between the coefficients in the effective
Hamiltonian (\ref{H}): $g_1=\xi(t_l)$ and $g_2=\xi(t_s)$, where
$\xi(t)$ is proportional to the PR electric field at
times $t_l$ and $t_s$, respectively. The quantity $c \cdot
(t_l-t_s)$ corresponds to the imbalance of the interferometers,
using the notation of equation (\ref{entanglement}). The previous
general condition $g_1=g_2$ of equation (\ref{general-condition})
therefore translates, in the case of up-conversion, as a
constraint on the PR's coherence: $\mathcal{L}_c^{PR}/c \gg
t_l-t_s$. In our experiment this condition is definitely satisfied
as $\mathcal{L}_c^{PR} > 1\,km$, which is clearly much greater
than the $20\,cm$ path length difference of Bob's analyzer.
Accordingly, the resulting photons at $\lambda_{B'}$ should become
entangled with Alice
photons at $\lambda_A$, although they never directly interacted.\\
\indent The PR delivers 700\,$mW$ in a standard telecom
fiber. We measured, with classical fields, an internal
up-conversion efficiency of 80\%/Watt of reservoir power at
optimum phase-matching\cite{Fejer04,Albota04,Vander-Up}. Note that
this value is underestimated since the losses in the waveguide
itself are neglected. This up-conversion efficiency decreases by a
factor of 2 when changing the single photon wavelength by $\pm
1\,nm$. Taking advantage of the energy correlation between the
photons of a pair, we reduced the bandwidth by filtering the
spectrum of the photons at 1555\,$nm$ travelling to Alice to
1.5\,$nm$ ($BPF_A$)\cite{Tittel-Weihs} and increased the power of
the source pump laser in order to have a reasonable photon-pair
production rate in this bandwidth. Hence, the overall
up-conversion probability is only limited by the available pump
power and coupling losses between the waveguide and single mode
optical fibers at input and output faces. From this classical
conversion efficiency, taking into account the reservoir power, a
realistic 40\% coupling efficiency into the waveguide (both for
the PR and the qubit to be transferred) and the ratio between the
initial and final wavelengths, we estimate the probability of a
successful QI transfer to be: $P_{success}=80\%\,W^{-1} \cdot
0.7\,W \cdot (0.4)^2 \cdot \frac{712\,nm}{1312\,nm} \approx 5\%$.
Let us mention that indirect qubit transfer can also be achieved
via teleportation\cite{MarcikicTele}. However, in contrast to
teleportation, QI transfer is achieved with a much higher success probability. \\
\indent After the up-conversion stage, we need to separate the
$\lambda_{B'}$ photons from the huge flow of PR photons at
1560\,$nm$, about $5\cdot10^9$ photons per $ns$, and measure the
transfer fidelity. For this, we use a bandpass filter centered at
712\,$nm$ ($BPF_B$, $\Delta \lambda = 10\,nm$, more than 30\,$dB$
attenuation around 1550 $nm$), as depicted in figure 3. We also
take advantage of the fact that the Si-APD ($D_B$) is essentially
insensitive to the 1560\,$nm$ wavelength. The 712.4\,$nm$ photons
then enter the temperature stabilized bulk-optic Michelson
interferometer. At the outport of this interferometer they are
coupled, for optimal mode overlap, into a single-mode fibre
adapted to visible light. The corresponding coupling efficiency is
greater than 60\%. Finally these photons are sent to the detector
$D_B$. The electronic signal from this APD triggers the detector
of the 1555\,$nm$ photons that arrive on Alice's side (InGaAs APD
previously introduced and operated in gated mode). Finally
we record the coincidence events.\\
\indent
Figure 3 shows the interference pattern of the
coincidence events as a function of the combined phase
($\phi_A + \phi_B$) obtained with our time-resolved coincidence
detection.
We observe a sinusoidal oscillation with net and raw visibilities of 96.2$\pm
0.4$\% and 86.4$\pm 0.4$\%, respectively, representing a clear
signature of the preserved coherence during the quantum
information transfer. The net visibility $V_{net}$ is again very
close to the 100\% predicted by quantum theory. Assuming that the
entire reduction of this visibility can be attributed to an
imperfect quantum state transfer, we find a fidelity $F =
\dfrac{1+V_{net}}{2}$ of 98.5\%. Note however that this estimation
is very conservative since the visibility observed before transfer
(see figure 2) is hardly
higher.\\
\indent Note that the quantum interface demonstrated in this
letter is not limited to the specific wavelengths chosen. Indeed,
suitable modifications of phase-matching conditions and pump
wavelengths enable tuning to any desired wavelengths. In
particular it allows mapping of telecommunication photonic qubits
onto qubits encoded at a wavelength corresponding to alkaline
atomic transitions. It is particularly interesting to take
advantage of periodically poled waveguiding structures, as
employed here. Firstly, phase matching conditions can easily be
tuned over a broad range by changing the grating period. Secondly,
these components yield very high up and down-conversion
efficiencies\cite{Tanz02,Bana01,Fejer04}. This permits the use of
a modest reservoir power to achieve a reasonable qubit transfer
probability. In the case of applications requiring very narrow
photon bandwidths, for instance when transferring quantum
information onto atoms\cite{Shapiro}, bright down-converters make
very narrow spectral filtering possible while maintaining high
photon-pair creation rates with reasonable pump powers. In
addition, the narrow filtering ensures optimal phase-matching over
the whole bandwidth of the photons to be up-converted, hence
maximizing this process.

%
\indent In this work we demonstrated, in the most general way, the
first direct quantum information interface between qubits carried
by photons of widely different wavelengths. To this end, we
verified that entanglement remains unaffected, even though one of
the two entangled photons is submitted to a wavelength
up-conversion process. This interface may find applications in
quantum networks, where mapping of travelling qubits, e.g. qubits
encoded onto telecommunication photons, and stationary atomic
qubits featuring resonance at much shorter wavelengths, is
necessary.

\vspace{2cm}
\noindent
We thank D.B. Ostrowsky for valuable discussions.\\
Financial supports by the Swiss National Center for Quantum Photonics and the European IST project
RamboQ are acknowledged.\\
S.T. acknowledges financial support from the European Science Foundation program "Quantum Information Theory and Quantum Computation".\\
The authors declare that they have no competing financial interests.\\
Correspondence and requests for materials should be addressed to S.T.


\begin{figure}[ht]
\centerline{\scalebox{.8}{\includegraphics{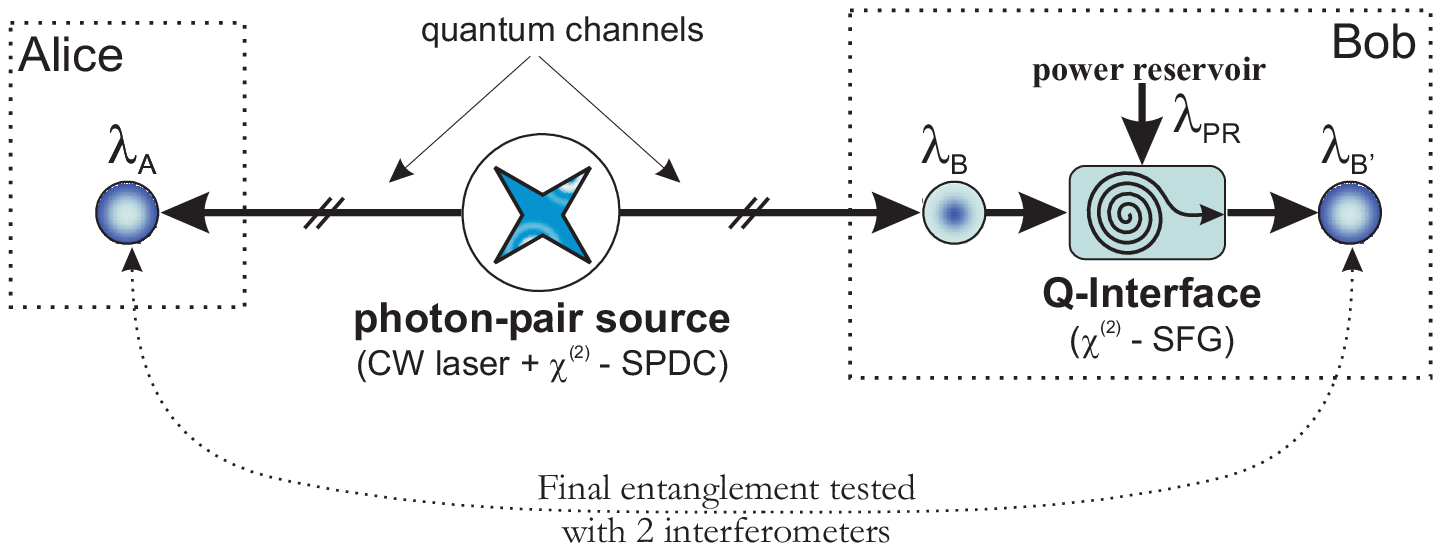}}}
\caption{
Schematic illustration of the experiment concept.
First, a photon-pair source produces, by spontaneous
parametric down-conversion (SPDC), energy-time entangled photons
at wavelengths $\lambda_A$ and $\lambda_B$ that are sent to Alice
and Bob, respectively. Next, the qubit transfer
is performed at Bob's place from photon $\lambda_B$ to
photon $\lambda_{B'}$ using sum frequency generation (SFG).
The final entanglement between the newly created photon $\lambda_{B'}$
and Alice's photon $\lambda_A$ is tested using the Franson
configuration\cite{Franson89} (see figure 2).}
\label{fig1ab-synoptic}
\end{figure}


\begin{figure}[ht]
\centerline{\scalebox{.8}{\includegraphics{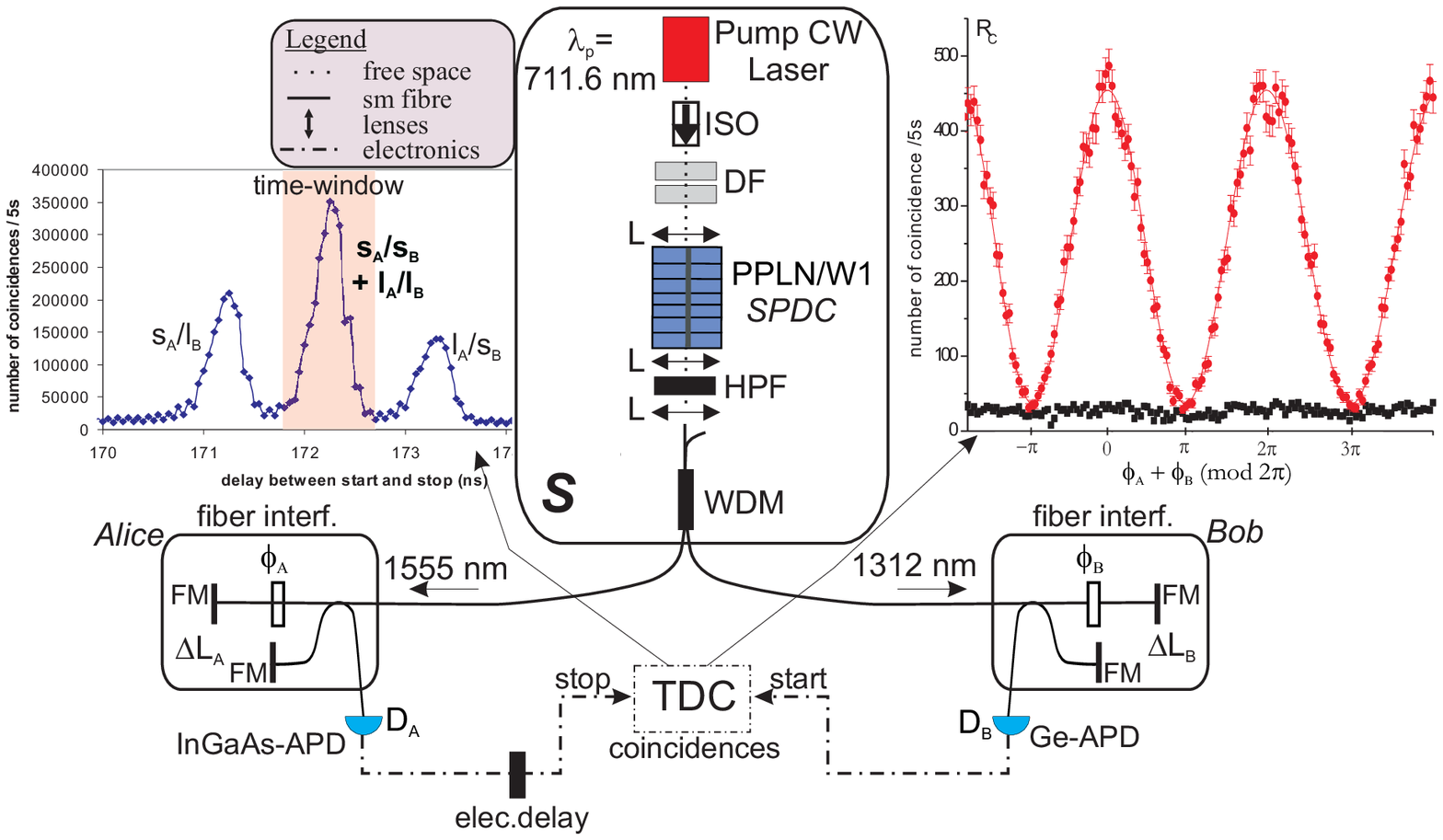}}}
\caption{
Experimental Franson-type setup used for the creation
and analysis of energy-time entangled pairs of photons.
Alice and Bob's analyzers are equally unbalanced Michelson interferometers.\\
The source is composed of a CW laser (Toptica Photonics DL100), a
nonlinear PPLN waveguide, a WDM and suitable fiber coupling
lenses. It also includes a high-pass wavelength filter
($HPF$) behind the crystal to discard the remaining pump photons,
and lenses ($L$) to couple the created photons into a single mode
fibre. The down-conversion quasi-phase matching is obtained with a
specific poling period $\Lambda$ of 14.1\,$\mu m$ and at a
temperature of 85$^{o}C$. This 1\,$cm$ long waveguide features a
down-conversion efficiency greater than 10$^{-7}$. At Bob's place the Ge-APD
(NEC, $D_B$) is liquid Nitrogen cooled and passively quenched,
while the InGaAs-APD (id Quantique id200, $D_A$) at Alice's is
operated in gated mode. The APDs show quantum detection efficiencies
of about 10\% and 14\% and probabilities of dark
counts of around $3 \cdot 10^{-5}$ and $10^{-5}$ per nanosecond, respectively.\\
The outputs from these APDs provide the start and stop signals for a
time-to-digital converter (TDC) that records a coincidence
histogram (left hand side) as a function of the time difference
between its two inputs. This picture is composed of three
different peaks which arise from different combinations of photon
transmissions trough their respective interferometer, either the
short ($s$) or the long ($l$) arm. The events where both photons
take the same arms ($s_A/s_B$ and $l_A/l_B$) are
indistinguishable as described by equation (\ref{entanglement}),
leading to photon-pair interference. They can
be discriminated from the other possibilities ($l_A/s_B$ and
$s_A/l_B$) by means of a time-resolved coincidence detection. The
corresponding coincidence count rate $R_C$ as a function of the
combined phase ($\phi_A + \phi_B$) is shown in the right hand
inset. We find sinusoidal interference fringes with more than 97\%
net visibility. Note finally that the variation of the combined
phase is obtained by changing the temperature in Bob's
interferometer.}
\label{franson-setup}
\end{figure}


\begin{figure}[ht]
\centerline{\scalebox{.8}{\includegraphics{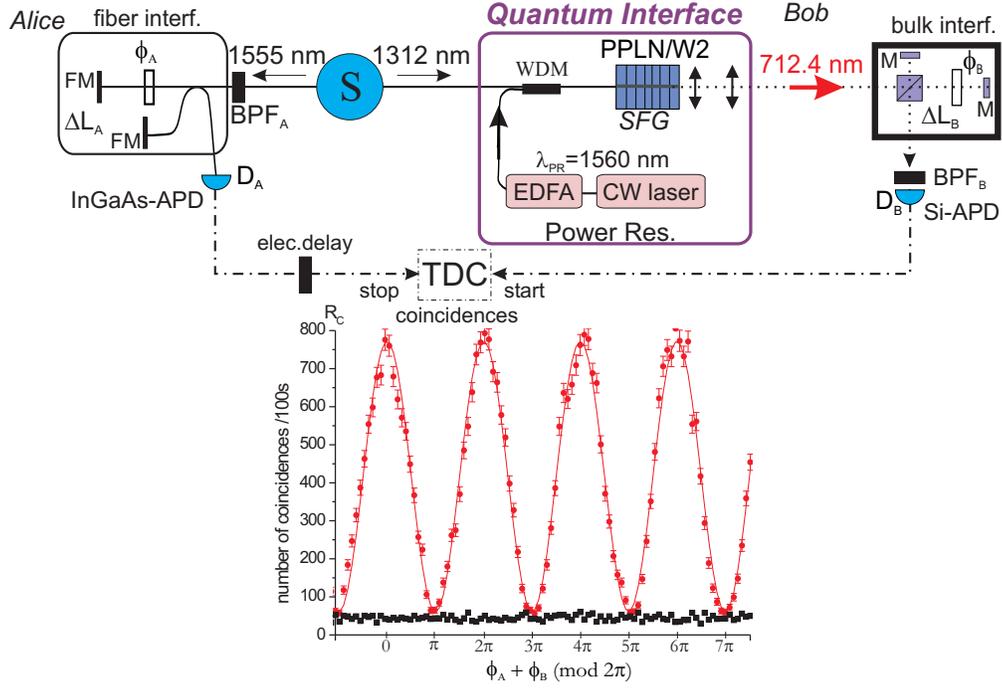}}}
\caption{
Experimental setup for the coherent transfer of quantum
entanglement. On Bob's side, PPLN/W2, pumped by a high-coherence
CW 700\,$mW$ power reservoir at 1560\,$nm$ (laser 8168A from HP +
EDFA from Keopsys), ensures the transfer of the qubit from 1312 to
712.4\,$nm$ via the up-conversion process. This second 1\,$cm$
long PPLN waveguide is of the same kind as the one used for the
down-conversion. They were both fabricated using the technique of
soft proton exchange\cite{Tanz02} and feature almost identical
phase-matching conditions.
Both 1312 and 1560\,$nm$ wavelengths are mixed at a second WDM
whose output port is directly butt-coupled to the input face of PPLN/W2
without any additional optics.
At the output of the waveguide, the newly
created photons at 712.4\,$nm$ are coupled into a single mode
fiber and detected using a Silicon
avalanche photodiode (EG\&G AQ-141-FC, $D_B$).\\
The coherence of the transfer is verified by measuring photon-pair
interference between the 712.4 and 1555\,$nm$ photons in the
Franson configuration. The analysis on Bob's side utilizes an
unbalanced, bulk-optic interferometer that is aligned with respect
to Alice's. Although the coincidence rate is substantially reduced
due to a limited up-conversion probability and losses in the
interferometers, the analysis of the coincidence events
yields interference fringes with net and raw visibilities
of 96 and 86\%, respectively.
This confirms the previously obtained result without the quantum
interface and demonstrates, in the most general way, coherent
transfer of quantum information between two photons.}
\label{transfer-setup}
\end{figure}

\vspace{-29cm}

\end{document}